\documentstyle[epsfig]{ioplppt}          % use this for journal style
\newcommand{\JETSET}{\mbox{\sc JETSET}}
\newcommand{\PYTHIA}{\mbox{\sc PYTHIA}}
\newcommand{\HERWIG}{\mbox{\sc HERWIG}}
\newcommand{\ARIADNE}{\mbox{\sc ARIADNE}}

\newcommand{\LEPONE}{\mbox{\sc LEP 1}}
\newcommand{\LEPTWO}{\mbox{\sc LEP 2}}
\newcommand{\QCD}{\mbox{\sc QCD}}

\def\lapproxeq {\mbox{{\lower .7ex\hbox{$\;\stackrel{\textstyle
                  <}{\sim}\;$}}}}
\def\gapproxeq  {\mbox{{\lower .7ex\hbox{$\;\stackrel{\textstyle
                  >}{\sim}\;$}}}}
\def\cpc#1#2#3{19#3 {\em Comp.\ Phys.\ Commun.}~{\bf#1} #2}

\def\pl#1#2#3{19#3 {\em Phys.\ Lett.}~{\bf#1B} #2}
\def\pr#1#2#3{19#3 {\em Phys.\ Rev.}~D {\bf#1} #2}

\def\zp#1#2#3{19#3 {\em Z.\ Phys.}~C {\bf#1} #2}
\newcommand{\nchqqqq}{\mbox{$N_{\mathrm{ch}}^{\mathrm{4q}}$}}
\newcommand{\nchqqlv}{\mbox{$N_{\mathrm{ch}}^{\mathrm{qq\ell\nu}}$}}
\newcommand{\mnchqqqq}{\mbox{$\langle\nchqqqq\rangle$}}
\newcommand{\mnchqqlv}{\mbox{$\langle\nchqqlv\rangle$}}
\newcommand{\WW}{\mbox{$\mathrm{W^+W^-}$}}
\newcommand{\qq}{\mbox{$\mathrm{q\overline{q}}$}}

\newcommand{\lv}{\mbox{$\ell\overline{\nu}_{\ell}$}}
\newcommand{\WWqqqq}{\mbox{\WW$\rightarrow$\qq\qq}}
\newcommand{\WWqqlv}{\mbox{\WW$\rightarrow$\qq\lv}}
\newcommand{\MW}{\mbox{$M_{\mathrm{W}}$}}
\newcommand{\Zz}{\mbox{${\mathrm{Z}^0}$}}
\newcommand{\Zqq}{\mbox{$\Zz/\gamma\rightarrow\qq$}}
\newcommand{\delnch}{\mbox{$\Delta\langle\nchqqqq\rangle$}}
\newcommand{\dndy}{\mbox{${\rm d}N_{ch}/{\rm d}y$}}
\newcommand{\epem}{\mbox{$\mathrm{e^+e^-}$}}
\newcommand{\kt}{\mbox{$k_{\perp}$}}
\newcommand\new{\newcommand}         % shorthand for \newcommand
\catcode`\"=\active                  % change category for \"
\new"[1]{{\accent127#1}}             % permit to do umlauts as "a
\new\3{\ss }                         % replace \ss by \3
\new{\mm}[1]{{\mbox{\hspace{#1mm}}}} % create horizontal space

\new\Tab[1]{Table~\ref{tab:#1}}

\new\dmw{\mbox{$\Delta \MW$}}

\newcommand{\DELPHI}{\mbox{\sc DELPHI}}
\newcommand{\OPAL}{\mbox{\sc OPAL}}

\newcommand{\LEP}{\mbox{\sc LEP}}

\begin{document} 

%\begin{tabbing}
%\` OPAL-CR324          \\
%\` 26$^{\mathrm{th}}$ August 1997.   \\
%\end{tabbing}

\title{Experimental aspects of colour reconnection}
 
\author{M.F.\ Watson\dag\ and N.K.\ Watson\ddag}

\address{\dag\ CERN, European Organisation for Particle
               Physics, CH-1211 Geneva 23, Switzerland}
\address{\ddag\ School of Physics and Space Research,
                University of Birmingham, Birmingham B15 2TT, UK}
 
\begin{abstract} 
  This report summarises experimental aspects of the phenomena of
  colour reconnection in \WW\ production, concentrating on charged
  multiplicity and event shapes, which were carried out as part of the
  ``Phenomenology Workshop on LEP2 Physics, Oxford, Physics Department
  and Keble College'', 14--18 April, 1997.  The work includes new
  estimates of the systematic uncertainty which may be attributed to
  colour reconnection effects in experimental measurements of \MW.
\end{abstract} 
 
% 
%  Uncomment out if preprint format required 
% 
%\pacs{00.00, 20.00, 42.10} 
\maketitle

\section{Introduction}
Colour reconnection (also referred to as `rearrangement' or `recoupling')
in \WW\ decays has been the subject of many studies (e.g.\ 
\cite{GPZ,SK,YB}) and at present there is agreement that observable effects
of interference between the colour singlets in the perturbative phase are
expected to be small.  In contrast, significant interference in the
hadronisation process appears a viable prospect but, with our current lack
of knowledge of non-perturbative \QCD, such interference can only be
estimated in the context of specific models
\cite{SK,GH,ARIADNE,HERWIG,NOVA,EG}. In the studies described below,
experimentally accessible features of these models\footnote{In studying
  these models, no retuning was performed when reconnection was enabled.}
are investigated, paying particular attention to the bias introduced to a
typical measurement of \MW\ by direct reconstruction of the decay products.

Throughout this section reconnection effects were studied using:
\PYTHIA~5.722, type I and type II superconductor models (with the string
overlap integral in type I case characterised by $\rho=0.9$) \cite{SK,YB};
\ARIADNE~4.08 allowing reconnection between the two W bosons; and
\HERWIG~5.9, in both its default reconnection model and also a `colour
octet' variant in which merging of partons to form clusters was performed on
a nearest neighbour basis\footnote{This was suggested by B~R~Webber, as a
  partial emulation of the model of reference \cite{EG}.}. In all cases, the
tuning of the models was as used in reference \cite{OPAL161}.

\section{Inclusive charged multiplicity}

It has been suggested \cite{SK,GH} that simple observable quantities such
as the charged multiplicity in restricted rapidity intervals may be
sensitive to the effects of colour reconnection. More recently \cite{EG} it
was suggested that the effect on the inclusive charged multiplicity itself
may be larger than previously considered and that the mean hadronic
multiplicity in \WWqqqq\ events, \mnchqqqq, may be as much as 10\% smaller
than twice the hadronic multiplicity in \WWqqlv\ events, \mnchqqlv. It was
also reported during this workshop that the effects of Bose-Einstein
correlations may increase \mnchqqqq\ by $\sim 3$--$10$\% 
(see \cite{bose-eins}).

The shifts in \mnchqqqq\ at the hadron level predicted by the models
studied thus far are given in table~\ref{tab:cr_nch}, where \delnch\ is
defined as the change in mean multiplicity relative to the `no
reconnection' scenario of each model.  From these, it is clear that the
multiplicities themselves and also the magnitude and sign of the predicted
shifts are model dependent.

\begin{table}
\caption{Mean charged multiplicities, \mnchqqqq,
 and predicted shifts for various models}
 \label{tab:cr_nch}
\begin{indented}
 \lineup
\item[]\begin{tabular}{@{}llll}
\br
model & & \mnchqqqq\ & \protect\delnch\ (\%) \\
\mr
 \PYTHIA\  &  normal      &  38.64 & \\
         &  type I      &  38.21 & $-1.1\pm$0.1 \\
         &  type II     &  38.39 & $-0.7\pm$0.1 \\
 \HERWIG\  &  normal      &  37.07 & \\
         &  reconnected ($P=\frac{1}{9}$) &  37.25  & +0.5$\pm$0.1 \\
         &  reconnected ($P=1$) &  38.38  & +3.5$\pm$0.1 \\
 \ARIADNE\ &  normal      &  38.14 &               \\
         &  reconnected &  37.07 &         $-2.8\pm$0.1 \\
\br
\end{tabular}
\end{indented}
\end{table}

In this study, the precision with which such tests may be performed is
quantified.  As a starting point for such tests, it was first verified that
in the absence of reconnection effects $\mnchqqqq=2\mnchqqlv$ in the models
\PYTHIA\ and \HERWIG. The statistical uncertainty of this test was ${\cal
  O}(0.1\%)$.  Next, samples of $10^5$ \HERWIG\ and \PYTHIA\ \WW\ events were
generated at $\sqrt{s}=171$~GeV with a full simulation of the \OPAL\
detector, and realistic event selections were applied for both \WWqqqq\ and
\WWqqlv\ ($\ell=$ e, $\mu$ and $\tau$).  The efficiency in each case was
$\sim$80\%, while the purity is $\sim 80\%$ for \WWqqqq\ and $\sim 88\%$
for the \WWqqlv\ channel.

\begin{figure}[tb]
  \epsfxsize=\textwidth
  \epsffile{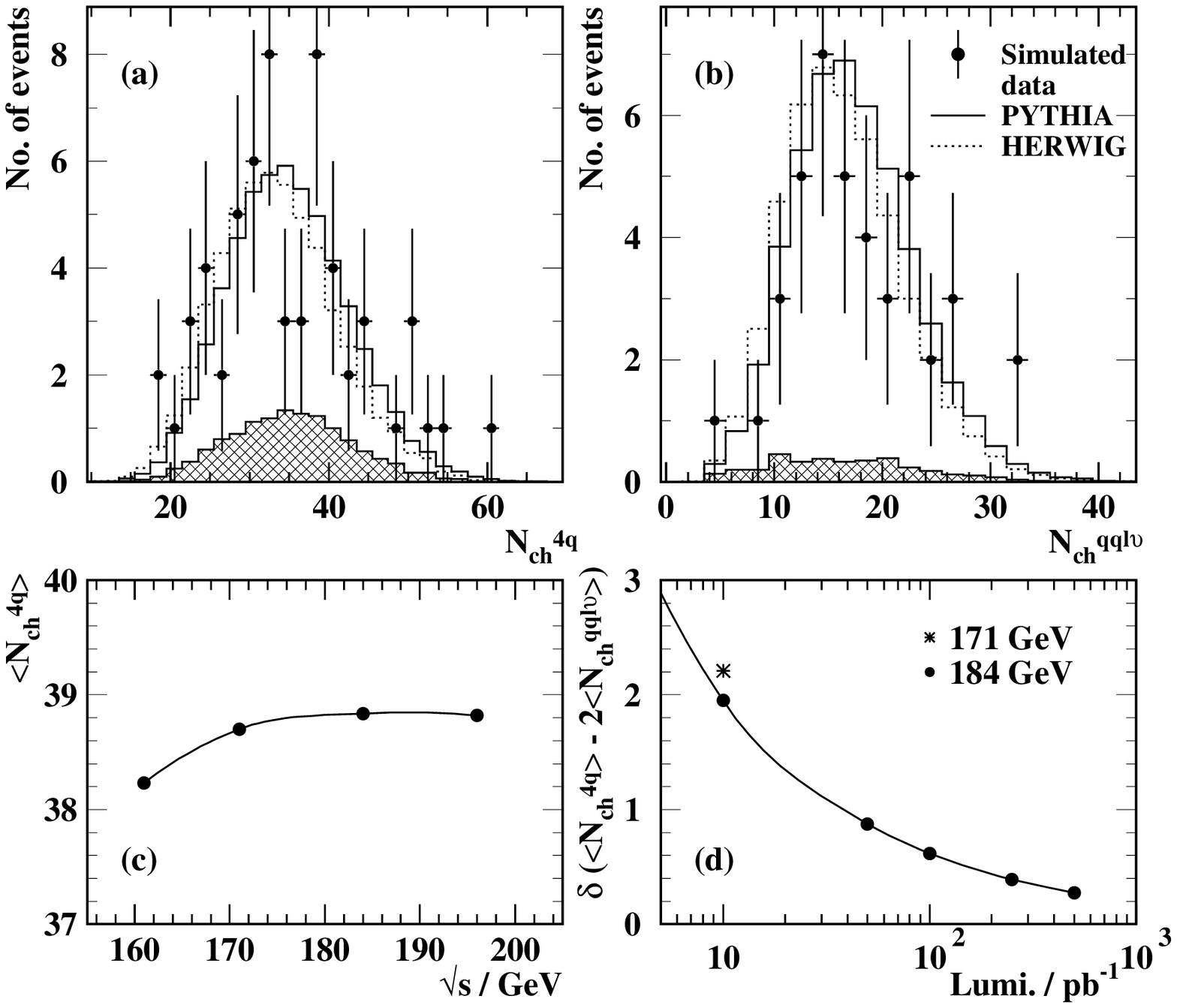}
 \caption{Inclusive charged multiplicity distributions with 10~pb$^{-1}$
   of fully simulated data, with background indicated hatched, at
   $\protect\sqrt{s}=171$~GeV for (a) \protect\WWqqqq, and (b)
   \protect\WWqqlv\ events.  (c) Variation of
   \protect\mnchqqqq\ with $\protect\sqrt{s}$.  (d)
   Luminosity dependence of the statistical uncertainty of
   $\protect\mnchqqqq-2\mnchqqlv$ (units of
   multiplicity).}
  \label{fig:cr_nch}
\end{figure}

The resulting (uncorrected) charged multiplicity distributions for the
hadronic and semi-leptonic channels are shown in Figs.~\ref{fig:cr_nch}(a)
and~\ref{fig:cr_nch}(b), respectively. The simulated data correspond
to an integrated luminosity of 10~pb$^{-1}$ at $\sqrt{s}=171$~GeV,
i.e. that delivered by \LEP\ in 1997. In both distributions, the expected
background is shown as a hatched histogram. The significant level of \Zqq\
background is apparent in the fully hadronic channel.

 To extract the mean charged multiplicity at the hadron level at a fixed
 centre-of-mass energy from such distributions, one can apply a simple
 correction, based on Monte Carlo, to the observed mean value, 
 after subtracting the expected
 background contribution. An alternative is to carry out a matrix-based
 unfolding procedure using the event-by-event correlation between the
 charged multiplicity at the hadron level and that observed in the detector
 after all analysis cuts have been performed. A separate correction for the
 effects of initial state radiation are necessary in this latter case.  A
 third alternative is to integrate the fragmentation function but this is
 not discussed here.
 
 Based on the the simulated data in Fig.~\ref{fig:cr_nch}(a) and (b), the
 expected statistical uncertainty on the difference
 $\mnchqqqq-2\mnchqqlv$ for an integrated
 luminosity of 10~pb$^{-1}$ is 2.2 units, or 5.7\% on
 \mnchqqqq. The evolution of the precision of such difference
 measurements with more data is estimated using the following assumptions.
 Firstly, the distributions of \nchqqqq\ and \nchqqlv\ are seen to be
 relatively insensitive to changes in centre-of-mass energy once away from
 the threshold region, as illustrated by the energy dependence of
 \mnchqqqq\ in Fig.~\ref{fig:cr_nch}(c). Therefore
 both the mean and the corresponding rms
 are assumed constant at their 184~GeV
 values. Secondly, above $\sqrt{s}=184$~GeV the \WW\ production
 cross-section is predicted to vary by less than 10\% in the region up to
 $\sqrt{s}<200$~GeV, and so a constant cross-section of 16~pb is assumed.
 Thirdly, it is assumed that the selection efficiency at 171~GeV may be
 maintained at higher energies. The expected background cross-section is
 not important as it is subtracted in performing the measurement.  Given
 these assumptions, the dependence of the expected statistical error on the
 difference, $\delta(\mnchqqqq-2\mnchqqlv)$, is
 shown as a function of integrated luminosity in Fig.~\ref{fig:cr_nch}(d).

 Typically in such multiplicity determinations, systematic effects become
 significant below a statistical precision of 0.5 units of multiplicity.
 Uncertainty in the modelling of 4-jet like \Zqq\ background with parton
 shower Monte Carlos in the
 fully hadronic channel may become a significant systematic.

\section{Event shapes}
Global event shape variables have been considered in earlier studies as
potential signatures for reconnection \cite{SK,GH,EG}. In most studies the
predicted effects on such observables induced by reconnection has been
sufficiently small that detection would be marginal,
even with an integrated luminosity of
500~pb$^{-1}$. 

The choice of a `no
reconnection' reference sample with which to compare data deserves some
thought.  In trying to find sensitive observables, using the models alone
is ideal. However, once possible signatures have been developed,
and one starts to search for effects in data, 
it will be invaluable
to have a well defined `no reconnection' reference sample in data 
to reduce model and tuning dependence.
\LEPONE\  data provide a high statistics reference, but additional assumptions
are necessary in either extrapolating energy scales, or in combining pairs
of \Zqq\ to emulate \WWqqqq\ events without reconnection.
It is also necessary to
assume that data recorded and processed by the detectors before 1996
can be directly compared with those recorded near the end of the \LEPTWO\
programme.  For some signatures, the ideal reference data are \WWqqlv\ 
events. However, this sample has only limited size and the comparison may
require the association of pairs of jets with Ws in the fully hadronic
channel, a procedure which experimentally introduces more uncertainty.  In
the following, all changes are relative to the `no reconnection' version of
each Monte Carlo model and all samples are \WWqqqq.

\begin{figure}[tb]
  \epsfxsize=\textwidth
  \epsffile{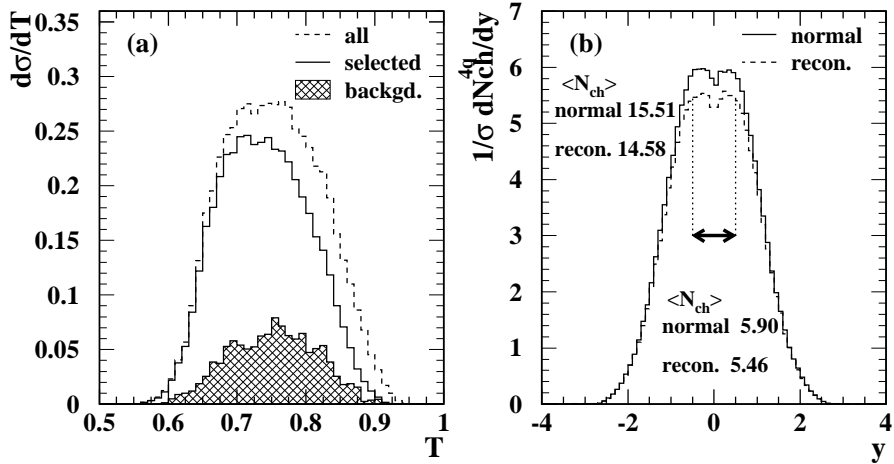}
 \caption{(a) Effect of typical experimental selection on thrust
          distribution, and (b) hadron level rapidity
          distribution in \ARIADNE\ for $p<1$~GeV.}
 \label{fig:cr_thrust_dndy}
\end{figure}

This study compares the differences in the rapidity distribution of charged
particles, \dndy, relative to the thrust axis of each event, in the central
region, $|y|<0.5$ and for all $y$, as suggested in \cite{GPZ,SK,GH}.  As
the effects are expected to be more pronounced for softer particles, the
distribution is studied for three momentum ranges, $p<0.5$~GeV, $p<1$~GeV
and all momenta.  It has been suggested \cite{GH,EG} that reconnection
effects may be more pronounced in specific topologies where the quarks from
different Ws are close to one another, therefore events are also studied
for all thrust values and for $T>0.76$.  One aspect not considered in
previous studies has been the effect of applying a realistic event
selection, which is necessary to reduce the large background ($\sigma(\Zqq)
\sim 20\sigma(\WWqqqq)$). As this is dominated by two-jet like events, the
efficiency for selecting \WWqqqq\ events in a similar configuration is
relatively small, as illustrated in Fig.~\ref{fig:cr_thrust_dndy}(a);
  $\sim38$\% of \WWqqqq\ events selected satisfy $T>0.76$, falling to
  $\lapproxeq 0.05$\% for $T>0.92$.

In \cite{EG}, the rapidity was studied relative to the axis bisecting the
two di-jet axes, as a function of the angle separating these axes.
Experimentally, without any reliable charge identification algorithm to
separate quarks from anti-quarks, the specific angle proposed in \cite{EG}
must at best be folded in experimental analyses, and also requires pairing
of jets into Ws. While the reliability of associating the `correct' jets
together is possible with moderate efficiency using kinematic fits,
selecting high thrust events was used in the current studies for expediency
and simplicity.  As the shifts in \MW\ expected are modest compared to the
experimental mass resolution on an event-by-event basis, it is worth
considering the use of kinematic fits in which our current knowledge of
\MW\ is applied as a constraint, in a similar way to that used by
experimental TGC analyses.

Hadronic events were generated using the models \PYTHIA, \HERWIG\ and \ARIADNE,
with and without a simulation of the \OPAL\ detector, and \dndy\ studied
within the ranges of $y$, $p$ and $T$ described above. A smearing
simulation of the \OPAL\ detector, which is reliable for studies in the
\WWqqqq\ channel and necessary to achieve the relatively high statistics
required, was used herein and also to estimate shifts in \MW.

\begin{figure}[tb]
%%% \centerline{\epsfig{file=dndy_trends.eps,width=\textwidth}}
 \epsfxsize=\textwidth
 \epsffile{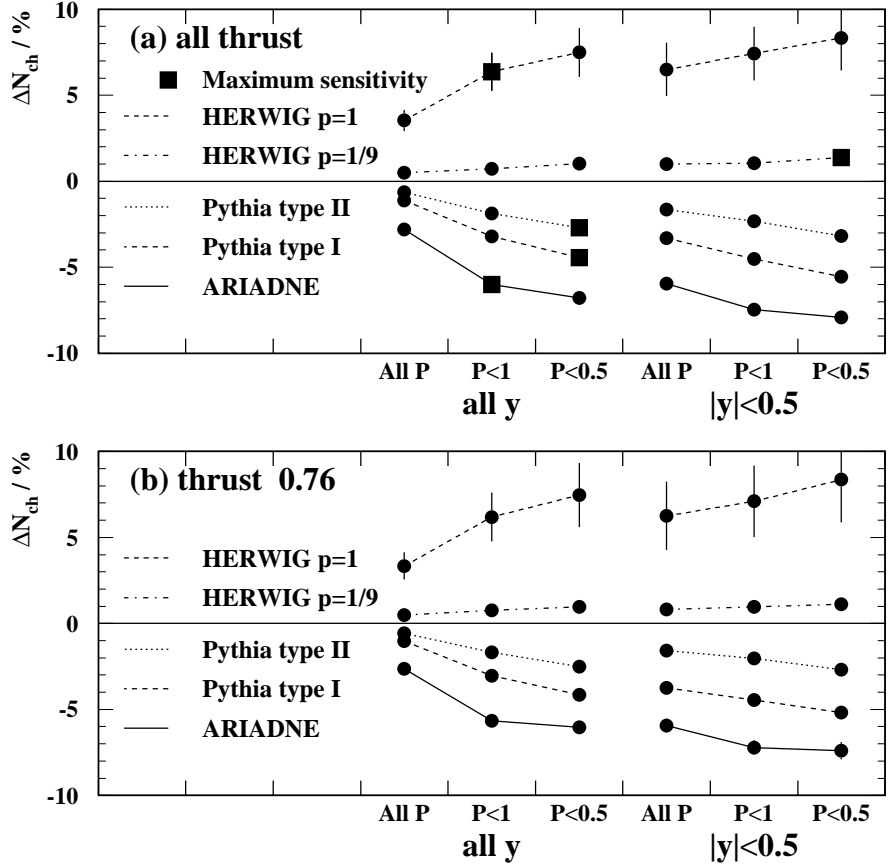}
   \caption{Fractional change in charged multiplicity as a function of
            maximum particle momentum, in two rapidity regions, for (a)
            all \protect{$T$}, and (b) \protect{$T>0.76$}. See text for
            details.}
   \label{fig:cr_dndy_trends}
\end{figure}

As an example of how the differences may be concentrated in restricted
rapidity intervals, Fig.~\ref{fig:cr_thrust_dndy}(b) shows the \dndy\ 
distribution for $p<1$~GeV in \ARIADNE, for events with and without
reconnection. Changes in charged multiplicity, \delnch, within given $p$
and $y$ intervals are summarised in Fig.~\ref{fig:cr_dndy_trends}(a) for
each of the models introduced in table~\ref{tab:cr_nch}, without detector
simulation.  The left (right) hand side of the figure shows the percentage
change in \mnchqqqq\ for the three momentum ranges considered
for all $y$ ($|y|<0.5$). The leftmost points in this figure correspond to
the results of table~\ref{tab:cr_nch}.  Fig.~\ref{fig:cr_dndy_trends}(b)
gives the analogous results for $T>0.76$.  For illustration, statistical
errors corresponding to an integrated luminosity of 500~pb$^{-1}$ are given
for the `\HERWIG\ colour octet' model.

It is seen that in all models the magnitude of the change increases when
only low momentum particles are considered. Applying a thrust cut such as
$T>0.76$ rejects $\sim40$\% of events and may change \mnchqqqq\ by up to
two units, but differences relative to the `no reconnection' scenarios are
essentially unchanged, therefore the sensitivity is reduced. The predicted
maximum statistical significance of \delnch, as well as its sign, depends
strongly on the model, varying from $\sim 6\sigma$ for \ARIADNE\ and the
\HERWIG\ `colour octet' model, $\sim 3.5\sigma$ for \PYTHIA\ type I, $\sim
2\sigma$ for \PYTHIA\ type II, down to $\sim0.8\sigma$ for \HERWIG. The
point of maximal sensitivity is indicated (square markers) for each model in
the figure. Similar trends were observed in studies with detector simulation
but typically \delnch\ was found to be $\sim$ 50\% smaller.

It may be possible to increase the sensitivity to reconnection effects
using charged multiplicity based methods, by considering particle
distributions relative to the \WW\ decay axis, as reconstructed using
kinematic fits. In \cite{YB}, an alternative multiplicity signature
(`interjet multiplicity') was introduced, having similar sensitivity to
integrating \dndy\ for $|y|<0.5$. This interjet multiplicity was similar in
idea to methods normally used to quantify the `string effect' in 3-jet
\epem\ events. It was suggested that this be studied further, using the
shape of the particle density distribution as a function of the angular
separation between jet pairs, rather than restricting the study to the
integrated particle density in the fixed angular regions. However, the
4-jet case is somewhat more complex than the familiar 3-jet case, being
non-planar, and so this was not pursued during the workshop.

\section{Shifts in \MW}

Extracting \MW\ from the decay products observed by experiments is
non-trivial, requiring much attention to bias induced from effects such as
initial state radiation, detector calibration, imperfect modelling of the
underlying physics processes and of the apparatus. In comparison to this,
estimating a shift which could result from the effects of reconnection
phenomena is straightforward, as the value of interest is the relative
shift between \MW\ determined in two different scenarios of the same
model. The absolute value of ``\MW'' obtained is not central to these
studies. However, there are still many uncertainties inherent in such
studies, such as sensitivity of the method used to extract \MW\ to changes
in $\sqrt{s}$, to tuning of the Monte Carlo models (e.g.  virtuality
cut-offs in the parton shower development), to treatment of combinatorial
background and ambiguous jet-jet combinations, and the range over which
fitting is performed to name but a few.

\begin{figure}[tb]
  \epsfxsize=\textwidth
  \epsffile{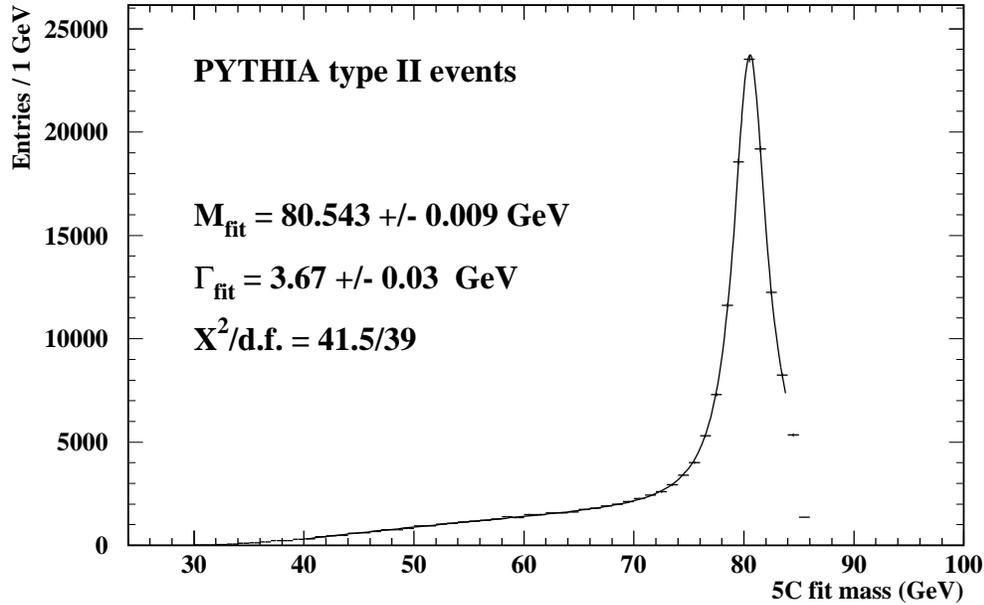}
  \caption{Typical mass distribution with fit results with
           detector simulation and event selection.}
  \label{fig:cr_mw_example}
\end{figure}

In these studies, the method used to extract \MW\ followed closely that
used by \OPAL\ for its preliminary \MW\ results using 172~GeV data. In this,
events with detector simulation are first selected using the same procedure
as noted earlier. Four jets are formed using the \kt\ jet finder, corrected
for double counting of energy within the apparatus, and a parametrisation
of the errors on the measured jet 4-momenta is carried out. A 
five-constraint kinematic fit, in which the jet-jet pair masses are constrained
to be equal, is performed for each of the three possible jet-jet pairings,
event by event. A mass distribution is constructed using
the mass from the combination having the highest probability from the
kinematic fit in each event if this has probability greater than 1\%. 
A second entry
is also admitted if the second most probable fit result has probability greater
than 1\% and within a factor of three of the highest probability combination.
The aim of this is to include additional mass information for events in
which the most probable fit combination is incorrect. In such events,
these two masses are essentially uncorrelated. A typical mass distribution
formed by this procedure is given in Fig.~\ref{fig:cr_mw_example}.

This method was applied to simulated events from each of the models in turn,
and the shifts obtained are summarised in table~\ref{tab:cr_deltamw}, where
uncertainties on these shifts are statistical.  The \ARIADNE\ model predicts
a modest shift in mass of approximately 50~MeV. No significant shift is
predicted by the models \PYTHIA\ and \HERWIG. In an earlier study, performed
in a similar way, significant shifts were determined \cite{YB}.  The
\PYTHIA\ and \ARIADNE\ models considered in the present study were also
included in \cite{YB}, albeit with different model dependent parameters and
looser event selection criteria.
\begin{table}
 \caption{Table of shifts in \MW\ for each model.}
 \label{tab:cr_deltamw}
 \begin{indented}
 \lineup
 \item[]\begin{tabular}{@{}llll}
 \br
       & & \multicolumn{2}{c}{$\protect\langle\Delta\MW\rangle$ (MeV)} \\
 model & & selected events ($\epsilon\simeq80$\%)  & all events \\
\mr
 \PYTHIA\  &  type I      &  $+18\pm$11 & $+11\pm$11 \\
         &  type II     &  $-13\pm$11 & $-19\pm$11 \\
 \HERWIG\  &  reconnected ($P=\frac{1}{9}$) & $-16\pm$16 & $-19\pm$16 \\
         &  reconnected ($P=1$)           & $+13\pm$15 & $+8\pm$14 \\
 \ARIADNE\ &  reconnected &  $+51\pm$16 & $+51\pm$15 \\
\br
\end{tabular}
\end{indented}
\end{table}

One quite plausible explanation proposed was that the difference was due to
the significantly more stringent event selection currently used. It has
been shown that the current selection preferentially rejects events having
two-jet like characteristics, which is where reconnection effects may be
expected to be prevalant. The rejection of these events does not appear to
be the reason for small mass shifts, as a similarly small effect is observed
when all events are selected, as seen in table~\ref{tab:cr_deltamw}.

Many possible sources for the difference were investigated in the context of
the \PYTHIA\ models. Neither changes in the tuning of \PYTHIA/\JETSET\ by
\OPAL\footnote{Among these, the cut-off parameter, $Q_0$, was increased
  from 1.0~GeV in the similar investigation of \cite{YB}, to 1.9~GeV.}  to
improve the description of \LEPONE\ data, nor the different centre-of-mass
energy ($\sqrt{s}=175$~GeV in \cite{YB}) were found to be significant.  The
current analysis procedure is slightly different to that in \cite{YB}.
However, significant shifts are still found when the current procedure is
applied to the same simulated events used in \cite{YB}.  Conversely,
applying the former procedure of \cite{YB} to the samples herein does not
induce a significantly larger mass shift.

One apparently significant effect was found to be the choice of mass
assigned to jets in performing kinematic fits. As discussed in \cite{YB},
this choice is not unique.  In the analysis of \cite{YB}, the hadronic jets
were assumed massless whereas in the current studies, the measured jet mass
was used.  Re-analysing the same simulated events of \cite{YB} but assigning
measured masses to the jets reduces the mass shifts estimated, e.g.\ shifts
quoted in \cite{YB} of $130\pm40$~MeV (type I) and $50\pm40$~MeV (type II)
become $70\pm40$~MeV and $30\pm40$~MeV, respectively. For comparison, a
sample of 200\,000 fully hadronic type I events were generated at
$\sqrt{s}=175$~GeV using {\em identical\/} model parameters and program
versions, and analysed using the procedure of \cite{YB}, also using measured
jet masses. This yielded an estimated shift of $46\pm16$~MeV.  It should be
noted that fluctations due to finite Monte Carlo statistics have to be
considered when comparing with the results of \cite{YB}, in which samples
sizes for the analoguous studies were 50\,000 events.

Comparing the results for mass shifts in table~\ref{tab:cr_deltamw}
with multiplicity shifts in table~\ref{tab:cr_nch}, it can be seen
that any relationship between them is model dependent. Furthermore,
relatively large shifts in the charged multiplicity do not necessarily
lead to a significant shift in \MW.

\section{Future}
The future for experimental studies of colour reconnection is quite open.
There is clear model dependence in signatures and mass shifts may be
smaller than earlier proposed \cite{YB}, although there are other
models available \cite{NOVA,EG} which were not tested in this study from
which different conclusions may be drawn. A necessary condition for a model
to be taken seriously is that it should describe the data, therefore
 tuning of models has to be addressed. With the current statistical
precision of \LEPTWO\ data, none of the models has been put to a stringent
test. The effect of background cannot be ignored in the \WWqqqq\ channel
as it proves difficult to remove. More sophisticated selections may
be developed, but typically these make use of non-trivial correlations
between observables, which may be poorly described by the models.
A particular concern is the description of parton shower Monte Carlos
to describe the hard, 4-jet like background which is selected.
The remaining point of note is that given the model dependence inherent
to such studies, it is most important to develop signatures which can
be tested taking the `no reconnection' scenario from data themselves.

\end{document}